\documentclass{article}
\usepackage{graphicx} % Required for inserting images
\usepackage{amsmath}
\usepackage{pgfplots}
\usepackage{xcolor}
\usepackage{verbatim}
\usepackage[parfill]{parskip}
\usepackage[hyphens]{url}
\usepackage [english]{babel}
\usepackage [autostyle, english = american]{csquotes}
\MakeOuterQuote{"}

\pgfplotsset{compat=newest}
\title{The Hyperdrive Protocol: An Automated Market Maker for Fixed and Variable Rates 
}
\author{Jonny Rhea \\ jonny@delv.tech \and Alex Towle \\ alex@delv.tech \and Mihai Cosma \\ mihai@delv.tech}
\date{April 2024}

\begin{document}

\maketitle

\begin{abstract}
\noindent
   Hyperdrive is a protocol designed to facilitate the trading of fixed and variable rate assets. The protocol's unique pricing model consolidates liquidity into a single pool which addresses the challenges of fragmented liquidity across terms, eliminates the need for rollovers, and allows terms to be issued on demand. Its design meaningfully improves trading efficiency, liquidity provisioning, and user experience over existing fixed and variable rate protocol models.
\end{abstract}

\section{Introduction\protect\footnote{This Whitepaper, the underlying code, and related documents regarding Hyperdrive refer to certain terms for convenience and ease of reading - such as "bonds," "shares," and related terms. These uses are made for colloquial reference, and do not refer to their official meanings in the financial industry or in any law, regulation, or rule; they are not intended to (and do not) have any particular legal or regulatory significance.}}

At its core, Hyperdrive is an automated market maker (AMM) protocol for fixed rates and variable rates; however, it has several unique features that set it apart from existing fixed and variable rate protocols:

\begin{itemize}
    \item \textit{Terms on Demand}: Hyperdrive allows for minting to be a part of the AMM, where the AMM essentially underwrites a new term for the user whenever they open a position. The user is not constrained to purchasing, selling, or minting into preexisting terms that are partially matured.
    \item \textit{Continuous Liquidity}: Hyperdrive pools never expire and underwrite a variety of fixed and variable rate terms with differing maturity dates. LPs can provide liquidity once without needing to roll their liquidity over to new terms.
    \item \textit{Single-Sided Liquidity}: Hyperdrive’s liquidity provision mechanism requires a single asset, improving both the UX and capital efficiency of LPing.
\end{itemize}

A Hyperdrive AMM allows users to open long and short positions to get exposure to fixed and variable rates, respectively. These rates are generated from trading activities and its base token, which is an arbitrary yield-bearing asset underlying the AMM. There are three position types in Hyperdrive:

\begin{enumerate}
    \item \textit{Long}: A long position represents some amount of fixed rate exposure to the market.  A user can purchase this with a base token and receive a bond token that will mature to the face value. Long traders receive fixed interest during the term and surrender any variable interest earned on their investment.
    \item \textit{Short}: When a user opens a short, they are selling bonds to the Hyperdrive pool - this increases the total supply of bonds available to purchase. Short traders sell long positions, which means they pay fixed interest during the term and receive all the variable interest collected on the LP's investment.
    \item \textit{Provide Liquidity}: LPs supply liquidity to a single pool, which underwrites a variety of fixed terms with differing maturity dates, giving LPs continuous exposure to the fixed rate market and underlying yield source. LPs, by definition, always take the opposite side of the trade opened by users.
\end{enumerate}

\section{Preliminaries}

% TODO: We should include some details on the pool's spot rate and spot price either here or in the pricing model. This would introduce $p$, which we use to describe fees in the position accounting section.
    
A Hyperdrive AMM calculates a market rate to provide users access to a fixed rate and variable rate. In typical AMMs, LPs deposit pairs of tokens into liquidity pools. Each pool consists of two different tokens--for example, ETH and DAI--and a function is then used to price the two assets in terms of each other for users to trade. One well known function for this, called the Constant Product Function \cite{Lu2017}, is defined as:

$$
\begin{aligned}
x \cdot y = k
\end{aligned}
$$

where the pair $(x,y)$ represents the pool reserves and $k$ is some fixed quantity that is preserved. The following shows how the function preserves $k$ when a user trades $\Delta x$ for $\Delta y$ with the pool:

$$
\begin{aligned}
(x + \Delta x) \cdot (y - \Delta y) = k
\end{aligned}
$$

Hyperdrive works similarly, except that instead of facilitating trades between ETH/DAI, it allows a user to buy and sell bonds that are backed by an underlying yield source.

\subsection{Yield Source}

A yield source allows a user to deposit a base token (x) and accrue positive yield over time. The rate of return of base tokens deposited into a yield source is not fixed and in many cases changes from block to block. This is referred to as the "variable interest rate" of a yield source protocol that holds deposited assets. 

Hyperdrive AMMs utilize Yield Sources to back the value of the bonds that are bought from and sold to it. When an LP provides liquidity to a Hyperdrive AMM, the assets are immediately deposited into a protocol that accrues variable rate interest. To support different yield sources without making large changes to the protocol, Hyperdrive implementations abstract away specifics of how yield source protocols accrue variable interest by introducing the concept of shares. 

\subsection{Shares}

The balance of assets held in a Hyperdrive pool is tracked in terms of “shares,” rather than base token amounts. Shares and base token amounts (i.e. "base") are related to each other in terms of a "share price," denoted as $c$, that changes over time. Given an amount of shares $z$, the corresponding amount of base $x$ can be calculated as:

$$
x = c \cdot z
$$

This share price abstraction provides a consistent way to measure interest accrual. Suppose that a Hyperdrive AMM's share reserves $z$ remain constant from time $t_0$ to time $t_1$. If $c_0$ is the share price at time $t_0$ and $c_1$ is the share price at time $t_1$, then the base reserves at $t_0$ are $x_0 = c_0 \cdot z$ and the base reserves at $t_1$ are $x_1 = c_1 \cdot z$. The amount of interest accrued between $t_0$ and $t_1$ is simply the difference in the base reserves $x_1 - x_0 = (c_1 - c_0) \cdot z$.  Note that for the remainder of this paper the variable, $c$, will represent the current "share price."

\section{Pricing Model}

Bonds in Hyperdrive are minted and given a maturity date upon purchase. This means the protocol must be able to price bonds with a variety of maturity dates from a single liquidity pool. The intuition behind how Hyperdrive accomplishes this is to re-imagine how a set of minted bonds matures. 
\pagebreak

Typical intuition would imagine all minted bonds maturing simultaneously: 
\begin{figure}[hbt!]
    \centering
    \includegraphics[width=.9\linewidth]{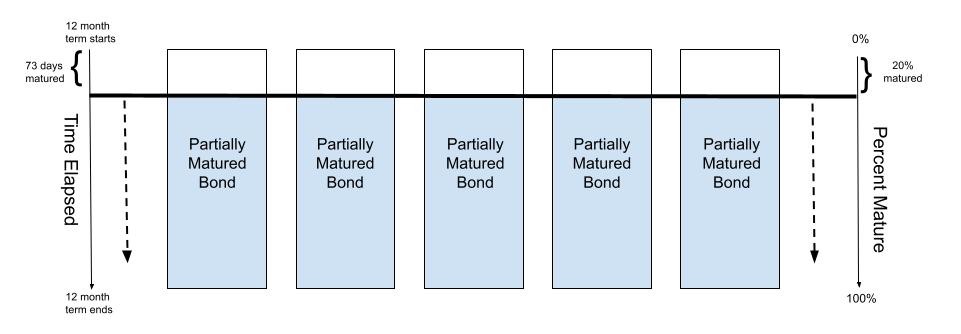}
    \caption{Bonds maturing simultaneously}
    \label{fig:bonds-maturing}
\end{figure}

In Hyperdrive, bonds mature sequentially:

\begin{figure}[hbt!]
    \centering
    \includegraphics[width=.75\linewidth]{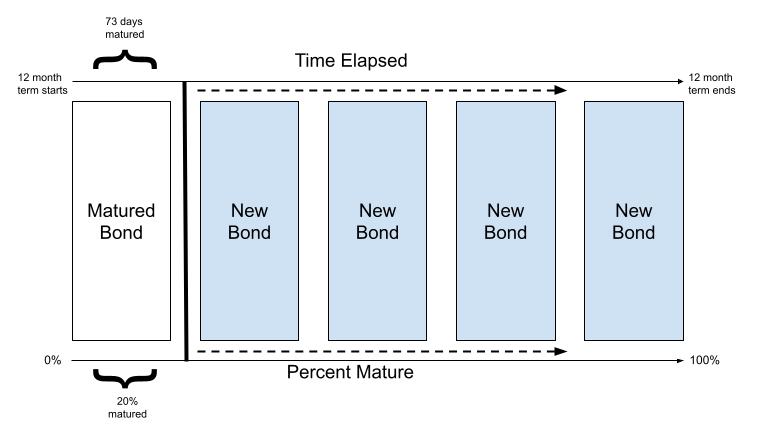}
    \caption{Bonds maturing sequentially}
    \label{fig:enter-label}
\end{figure}

This allows the AMM to price new bonds that haven't matured at the market value and matured bonds at their face value.

\subsection{Buying and Selling Bonds}

Hyperdrive trading is fundamentally comprised of two actions, buying and selling bonds. The order in which this is done defines the type of position the user is purchasing and how the reserves are impacted. Since bonds are minted on demand, they don't exist until a user opens a position that either buys them from the pool or sells them to the pool. Bonds in a Hyperdrive pool can be thought of as a virtual quantity representing how many bonds can be minted at the pool's current rate given the amount of share liquidity in the pool.

In this section (3.1), we will define some notation before addressing the impact that buying and selling bonds has on the reserves. Since the price of a new bond is driven by market demand, they are priced as a function of the share and bond reserve levels (z,y), respectively, using a trading invariant. We will refer to this trading invariant function as $I$.

$$
\begin{aligned}
\qquad \qquad \qquad \qquad \qquad \Delta out = I_{out}^{in} \left(\Delta in, z, y \right) \qquad \qquad \qquad \qquad (1)
\end{aligned}
$$

For the purposes of this discussion, the only requirement for how $I$ is defined is that it preserves path-independence. In other words, the end state of the reserves discussed will always be the same, regardless of the order in which trades are executed. As mentioned in the previous section, matured bonds are priced at their face value. It is important to note that, assuming the pool is solvent, this pricing is also path-independent.  We can define this pricing function as:

$$
\begin{aligned}
\qquad \qquad \qquad \qquad \qquad M \left(\Delta y \right) = \frac{\Delta y}{c} = \Delta z    \qquad \qquad \qquad \qquad \qquad(2)
\end{aligned}
$$

Combining (1) and (2) together, we define a function $H$ that can calculate a position's impact on $z$ given $\Delta y$ at time remaining $t_r$. We use the same "in" and "out" notation defined above for $I$. $H$ is defined as follows:

$$
\begin{aligned}
\qquad H_{\Delta z}^{\Delta y} \left(\Delta y, z, y, t_r \right) = I_{\Delta z}^{\Delta y} \left(\Delta y \cdot t_r, z, y \right) + M \left(\Delta y \cdot ( 1- t_r)\right) \qquad &&& (3)
\end{aligned}
$$

\subsubsection{Opening Positions}
 When opening a position, the pricing model uses the trading function $H$ to calculate the impact on the reserves at $t_r = 1$ and we always solve for the $\Delta z$ that keeps the pool's depth, $k$, constant. 

When a user opens a long, they are purchasing fixed-rate exposure at the current market rate. We can represent the impact that a purchase of $\Delta y$ bonds with $\Delta z$ shares has on the reserves as:

$$
\begin{aligned}
\Bigl(z + H_{\Delta z}^{\Delta y} \left(\Delta y, z, y, 1 \right)\,,\,y - \Delta y \Bigr)
\end{aligned}
$$

When a user opens a short, they are purchasing exposure to the underlying yield source's variable rate. In this case, the user is owed the amount of variable interest accrued over the course of the bond's term, meaning funds must be removed from the reserves to accrue interest without backing additional open positions. We can represent the sale of $\Delta y$ bonds for $\Delta z$ shares on the reserves as:

$$
\begin{aligned}
\Bigl(z - H_{\Delta z}^{\Delta y} \left(\Delta y, z, y, 1 \right)\,,\,y + \Delta y \Bigr)
\end{aligned}
$$

\subsubsection{Closing Positions}

When closing a position, the pricing model uses the time remaining, $t_r$, to determine the proportion of bonds that Hyperdrive considers new versus matured. As a result, when closing a position, we use $I$ to price the proportion of bonds we consider new and $M$ to price the matured ones. Since matured bonds can just be redeemed for the face value of the bond in terms of the base token, we know that at maturity $\Delta x = \Delta y$.

When a user closes a long, they are redeeming the $\Delta y \cdot (1-t_r)$ matured bonds and selling $\Delta y \cdot t_r$ new bonds back to Hyperdrive. This means that the reserves will be updated as follows:

$$
\begin{aligned}
\Bigl(z - H_{\Delta z}^{\Delta y} \left(\Delta y, z, y, t_r \right)\,,\,y + \Delta y \cdot t_r \Bigr)
\end{aligned}
$$

When a user closes a short, they are redeeming the $\Delta y \cdot (1-t_r)$ matured bonds and purchasing the $\Delta y \cdot t_r$ new bonds via Hyperdrive. Again, since the trader is buying the new bonds from Hyperdrive, the reserves will be updated as follows:

$$
\begin{aligned}
\Bigl(z + H_{\Delta z}^{\Delta y} \left(\Delta y, z, y, t_r \right)\,,\,y - \Delta y \cdot t_r \Bigr)
\end{aligned}
$$

Unfortunately, these accounting updates are problematic because the impact of $M$ on the share reserves changes the pool's depth; this allows a trader to open a position on one curve and close it on another by sandwiching a closing trade. As Peter Zeitz's Curve Vulnerability Report \cite{zeitz2020} demonstrated, it is not safe for an AMM to support several different trading curves simultaneously as it allows malicious traders to extract large amounts of value from the system. 

A modification to the accounting scheme is needed to hold the pool's depth constant before and after closing matured positions. In the next section, we will define such a scheme by introducing the zeta adjustment or $\zeta$.

\subsection{Zeta Adjustment}

Simply put, the zeta adjustment replaces the reserve tuple $(z, y)$ used when pricing new bonds and replaces it with $(z_e, y)$ where $z_e$ is defined as:

$$
\begin{aligned}
 \qquad \qquad \qquad \qquad \qquad \qquad z_e = z - \zeta \qquad \qquad \qquad \qquad \qquad \qquad (4)
\end{aligned}
$$

$\zeta$ is introduced as a new state variable to offset any changes to $z$ caused by $M$.  When a long is closed at time $t_r$, $z_e$ is calculated as follows:

$$
\begin{aligned}
z_e &= \biggl(z - H_{\Delta z}^{\Delta y} \left(\Delta y, z, y, t_r \right)\biggr) - \biggl(\zeta - M \bigl(\Delta y \cdot \left( 1 - t_r\right)\bigr)\biggr) \\
\implies z_e &= \left(z - \zeta\right) - I_{\Delta z}^{\Delta y} \left(\Delta y \cdot t_r, z, y \right)
\end{aligned}
$$

Similarly, when a short is closed at time $t_r$, $z_e$ is calculated as follows:
$$
\begin{aligned}
z_e &= \biggl(z - H_{\Delta z}^{\Delta y} \left(\Delta y, z, y, t_r \right)\biggr) - \biggl(\zeta + M \bigl(\Delta y \cdot \left( 1 - t_r\right)\bigr)\biggr) \\
\implies z_e &= \left(z - \zeta\right) + I_{\Delta z}^{\Delta y} \left(\Delta y \cdot t_r, z, y \right)
\end{aligned}
$$

As a result, when a position is closed at maturity, $z_e$ remains constant because the update to $z$ by $M$ is canceled out by the change made to $z_e$. Using the tuple $(z_e, y)$, rather than $(z, y)$, to parameterize the pricing model for new bonds will hold the spot price and depth constant.

The following plot illustrates the impact of the \textit{Zeta Adjustment}:

\begin{tikzpicture}
\begin{axis}[
    title={Zeta Adjustment: $(z,y)$},
    xlabel={$z$},
    ylabel={$y$},
    xmin=0, xmax=0.8,
    ymin=0, ymax=0.8,
    grid=both,
    legend pos=north east
]
% Plot z^(1-1/t) + y^(1-1/t) = k
\addplot[domain=-0.1:5, samples=400, unbounded coords=jump]{0.8 - x^(4/5))^(5/4)};
\addlegendentry{$\zeta=0$}

% Plot (z-\zeta)^(1-1/t) + y^(1-1/t) = k 
\addplot[domain=-0.1:5, samples=400, unbounded coords=jump, color=red]{0.8 - (x+0.1)^(4/5))^(5/4)};
\addlegendentry{$\zeta < 0$}

% Plot (z-\zeta)^(1-1/t) + y^(1-1/t) = k
\addplot[domain=-0.1:5.1, samples=400, unbounded coords=jump, color=blue]{0.8 - (x-0.1)^(4/5))^(5/4)};
\addlegendentry{$\zeta > 0$}

\addplot[domain=0:5, samples=400, dotted, unbounded coords=jump]{x};

\end{axis}
\end{tikzpicture}

When zeta is adjusted for a short close, it translates the curve to the left (red). When zeta is adjusted for a long close, it translates the curve to the right (blue). As mentioned previously, for the \textit{Zeta Adjustment} to cancel out the impact redemptions of matured bonds have on the pricing model, we must use $z_e$ in place of $z$ when calculating trades on $I$.  The following plot illustrates this:

\begin{tikzpicture}
\begin{axis}[
    title={Zeta Adjustment: $(z_e,y)$},
    xlabel={$z_e$},
    ylabel={$y$},
    xmin=0, xmax=0.8,
    ymin=0, ymax=0.8,
    grid=both,
    legend pos=north east
]
% Plot z^(1-1/t) + y^(1-1/t) = k
\addplot[domain=-0.1:5, samples=400, unbounded coords=jump]{0.8 - x^(4/5))^(5/4)};
\addlegendentry{$\zeta=0$}

% Plot (z-\zeta)^(1-1/t) + y^(1-1/t) = k 
\addplot[domain=-0.1:5, samples=400, unbounded coords=jump, color=red]{0.8 - (x)^(4/5))^(5/4)};
\addlegendentry{$\zeta < 0$}

% Plot (z-\zeta)^(1-1/t) + y^(1-1/t) = k
\addplot[domain=-0.1:5.1, samples=400, unbounded coords=jump, color=blue]{0.8 - (x)^(4/5))^(5/4)};
\addlegendentry{$\zeta > 0$}

\addplot[domain=0:5, samples=400, dotted, unbounded coords=jump]{x};

\end{axis}
\end{tikzpicture}

As seen above, all scenarios look identical from the $z_e$ axis. Since $I$ is assumed to be path-independent, $M$ is trivially path-independent and has no impact on $z_e$; this implies that the Hyperdrive pricing model as a whole is path-independent.

\section{Position Accounting}

Since Hyperdrive deposits the capital that underlies bonds into a yield source, interest accrues during the lifetimes of longs and shorts. In exchange for the fixed interest that longs receive as their bonds mature, they give up the variable interest that accrues over time. In contrast, shorts receive the variable interest that accrues over time to compensate them for paying fixed interest to the pool.

Fees are charged when traders open and close positions. LPs receive a fee proportional to the pool's spot rate when newly-minted bonds are traded that is parameterized by $\phi_n$, $0 \leq \phi_n \leq 1$. Additionally, LPs receive a fee proportional to position size when matured bonds are traded that is parameterized by $\phi_m$, $0 \leq \phi_m \leq 1$. Governance may collect a portion of trading fees parameterized by $\phi_g$, $0 \leq \phi_g \leq 1$. For convenience, we will define the following functions that represent the fee calculations:

$$
\begin{aligned}
f_{new}(\Delta y, t_r) &= \phi_n \cdot \left( 1 - p \right) \cdot \Delta y \cdot t_r\\
f_{mature}(\Delta y, t_r) &= \phi_m \cdot \Delta y \cdot \left( 1 - t_r \right)\\
f(\Delta y, t_r) &= f_{new}(\Delta y, t_r) + f_{mature}(\Delta y, t_r)
\end{aligned}
$$

These will be referenced as needed in the following sections.

\subsection{Longs}

Traders can open long positions to receive fixed-rate exposure. From the perspective of the pricing model, opening a long position is equivalent to purchasing bonds. This has the consequence of pushing the spot rate down and, conversely, pushing the spot price up towards the face value of the bond. Assuming the pool has sufficient liquidity, traders can close their existing long positions at any time before maturity, and regardless of the pool's idle liquidity (see Section 5 for more details on idle liquidity), they will always be able to close their long positions at maturity. 

\subsubsection{Opening Longs}

To open a long position, a trader provides $\Delta x$ base to a Hyperdrive AMM and receives a long position with a face value of $\Delta y$. The long's face value is equivalent to the amount of bonds underlying the long and the amount of base proceeds that can be redeemed when closing the bonds at maturity. Since longs are opened with a full term remaining before maturity, the size of long positions for newly minted bonds is priced using the invariant as follows:

$$
\Delta y = H_{\Delta y}^{\Delta z} \left( 
    \tfrac{\Delta x}{c}, z, y, 1
\right) - \phi_n \cdot \left( p^{-1} - 1 \right) \cdot \Delta x
$$

Since bonds are purchased when a trader opens a long, the share reserves are increased and the bond reserves are decreased. The purchased bonds are all newly minted, so the zeta adjustment isn't changed when updating the reserves. With all of this in mind, the pool's state is updated as follows:

$$
\begin{aligned}
    z_1 & = z_0 + \frac{
        \Delta x - \phi_g \cdot \phi_n \cdot \left( 1 - p \right) \cdot \Delta x
    }{c} \\
    y_1 & = y_0 - \Delta y \\
    \zeta_1 & = \zeta_0
\end{aligned}
$$

\subsubsection{Closing Longs}

When a trader closes a long position, they sell the bonds that underlie the position back to the pool and receive the base proceeds that were released from the sale. If the position is closed at or after maturity, the trader receives the face value of the bonds, and the pool's spot rate remains constant. If the position is closed before maturity, some of the bonds underlying the long are newly minted, and the pool's spot rate will increase as a result of the sale.

The base proceeds $\Delta x$ derived from closing a long position of size $\Delta y$ depend on the time remaining until maturity of $t_r$, $0 \leq t_r \leq 1$. We decompose the proceeds $\Delta x$ into the proceeds of selling newly minted bonds $\Delta x_0$ and the proceeds of selling matured bonds $\Delta x_1$. We compute the base proceeds $\Delta x$ as follows:

$$
\begin{aligned}
\Delta x & = c \cdot H_{\Delta z}^{\Delta y} \left( \Delta y, z, y, t_r \right) - 
   f(\Delta y, t_r)
\end{aligned}
$$

Since the trader sells bonds when they close a long, the share reserves are decreased by the amount of base proceeds the trader receives. The bond reserves are increased by the amount of newly minted bonds that were sold. To ensure that the pricing model is only affected by the sale of newly minted bonds, the zeta adjustment is decreased by the proceeds of the sale of matured bonds. The state updates resulting from closing the long are given by:

$$
\begin{aligned}
    z_1 & = z_0 - \frac{\Delta x + \phi_g \cdot f(\Delta y, t_r)}{c} \\
    y_1 & = y_0 + \Delta y \\
    \zeta_1 & = \zeta_0 - M(\Delta y \cdot (1 - t_r)) + \left( 1 - \phi_g \right) \cdot f_{mature}(\Delta y, t_r)
\end{aligned}
$$

\subsection{Shorts}

With respect to pricing, short positions are the inverse of long positions. They receive variable-rate exposure in exchange for fixed-rate exposure. From the perspective of the pricing model, opening a short is equivalent to selling bonds to the LPs. This has the consequence of decreasing the spot rate and increasing the spot price, which is the opposite of what happens when longs are opened. For the system to remain solvent, shorts must deposit the maximum amount of money that they can lose when the short is opened, which is equal to the fixed interest that will be owed to LPs at maturity. 

When a trader closes their short, they purchase back the partially (or fully) matured bond that underlies the short position at its current market price. By locking up only the fixed interest, they’ll earn the current market discount plus variable interest accrued by the full face value of the bonds. This gives short positions multiplied exposure to the variable rate.

\subsubsection{Opening Shorts}

To open a short position, a trader specifies the amount of bonds to short $\Delta y$. The short is responsible for depositing the difference between the value of the bonds plus any interest that has accrued since the start of the checkpoint (see section 4.3 for more details on checkpoints), $\tfrac{c}{c_0} \cdot \Delta y$, and the current price of the bonds. The short deposit is calculated as:

$$
\Delta x = \tfrac{c}{c_0} \cdot \Delta y - c \cdot H_{\Delta z}^{\Delta y}(\Delta y, z, y, 1) + f_{new}(\Delta y, 1)
$$

Since bonds are sold when a trader opens a short, the share reserves are decreased by the proceeds from selling the short, and the bond reserves are increased by the amount of bonds that were short-sold. The bonds that are short-sold are newly minted, so the zeta adjustment remains constant. With this in mind, the pool's state is updated as follows:

$$
\begin{aligned}
    z_1 & = z_0 - H_{\Delta z}^{\Delta y}(\Delta y, z, y, 1) + f_{new}(\Delta y, 1) \cdot ( 1  - \phi_g ) \\
    y_1 & = y_0 + \Delta y \\
    \zeta_1 &= \zeta_0
\end{aligned}
$$

\subsubsection{Closing Shorts}

When a trader closes a short position, they purchase the bonds that were short-sold when the short was opened. The proceeds that they receive are the difference between the face value of the underlying bonds $\Delta y$ and the cost to purchase the bonds plus the variable interest accrued since the short was opened. Let $c_0$ be the share price when the short was opened, and $c_1$ be equal to the current share price, $c$, if the shorts haven't matured; otherwise, $c_1$ is the share price at the time the shorts were closed. We can calculate the shorts trading proceeds as:

$$
\Delta x = \frac{c_1}{c_0} \cdot \Delta y - c \cdot H_{\Delta z}^{\Delta y}(\Delta y, z, y, t_r ) - f(\Delta y, t_r)
$$

Since the bonds that were short-sold are purchased when the short is closed, the share reserves are increased by the cost of purchasing the bond, and the bond reserves are decreased by the amount of bonds that were purchased. To ensure the pricing model is only impacted by the sale of newly minted bonds, the zeta adjustment is increased by the cost of the purchase of matured bonds. The state updates resulting from closing the short are given by:

$$
\begin{aligned}
    z_1 & = z_0 + H_{\Delta z}^{\Delta y} \left(\Delta y, z, y, t_r \right) + 
             \frac{f(\Delta y, t_r)}{c} \cdot  \left( 1 - \phi_g \right)   \\
    y_1 & = y_0 - \Delta y \\
    \zeta_1 & = \zeta_0 + M(\Delta y \cdot (1 - t_r)) + \frac{f_{mature}(\Delta y, t_r)}{c} \cdot (1 - \phi_g)
\end{aligned}
$$

\subsection{Checkpoints}

% TODO: Explain how time is segmented by `calculateTimeRemaining`.

Closing positions at maturity automatically increases the system's capital efficiency and ensures that positions aren't open for longer than the predetermined position duration. While closing positions individually would be theoretically possible, the costs would be prohibitive. With this in mind, positions are grouped into checkpoints. Each Hyperdrive pool is parameterized with a checkpoint duration $d_c$ that determines the size of these checkpoints. The checkpoint duration evenly divides the position duration. Given the current block time $t$, the start of the latest checkpoint $t_c$ is calculated as:

$$
t_c = t - (t \bmod d_c)
$$

When positions are opened, they are backdated to the start of the latest checkpoint. This ensures that all positions opened during a checkpoint mature simultaneously. The first trade or LP operation in a checkpoint mints the checkpoint by storing the current share price as the checkpoint's share price, which is used to calculate the proceeds of shorts opened in the checkpoint. Aside from storing the checkpoint's share price, minting the checkpoint also applies state updates resulting from closing matured positions in the checkpoint to the pool's reserves. Since the closed positions have already matured, the share reserves and zeta adjustment are updated by the amount of longs that have matured $y_l$ and amount of shorts that have matured $y_s$ as:

$$
\begin{aligned}
    z & \mathrel{+}= y_l - y_s \\
    \zeta & \mathrel{+}= y_l - y_s
\end{aligned}
$$

Traders could hypothetically wait until the end of the checkpoint to open long positions, allowing them to accrue fixed interest as soon as the checkpoint advances. To mitigate this arbitrage from negatively impacting LPs, the checkpoint duration should always be short enough that the fixed interest accrued is less than the cost of a trade. LPs should verify that the checkpoint duration is sufficiently small before interacting with a Hyperdrive pool.

\subsection{Solvency}

Since longs and shorts are entitled to fixed-rate and variable-rate exposure over the duration of the term, some of a Hyperdrive pool's capital must be reserved to ensure the system is able to honor this exposure at maturity. When shorts are opened, all of the base required to capitalize the shorts is reserved outside of the share reserves, which ensures that the capital underlying shorts can't be consumed by trading or LP actions. In contrast, the capital that underlies longs is initially held in the share reserves and can be consumed by trading and LP actions. This means that the system must enforce a solvency constraint on longs to prevent trading and LPs from removing the capital reserved for longs.

At maturity, longs receive the face value of their position, and shorts receive the interest generated on the face value of their position. Since longs receive the principal and shorts receive the interest, one base asset is sufficient to capitalize both a long and a short. This means that longs and shorts can net out. Since shorts already reserve capital equal to their face value, matching longs and shorts opened in the same checkpoint can use the same underlying collateral. The solvency constraint $e_c$ required to ensure that positions in checkpoint $c$ are solvent is calculated using the number of longs in checkpoint $y_l$ and shorts in checkpoint $y_s$:

$$
e_c = \max(y_l - y_s, 0)
$$

Since longs and shorts in different checkpoints mature at different times, these positions cannot be matched across checkpoints without compromising the fixed term of longs and shorts. So while positions within a single checkpoint can net out, the long exposure across checkpoints must be added up to determine the true solvency requirement. Assuming the number of checkpoints per term is $n \geq 1$ and the current checkpoint is $c$, the global solvency requirement for positions can be calculated as:

$$
e = \sum_{k = c - (n - 1)}^{c} e_k
$$

Since funds for non-netted longs are contained in the share reserves, the solvency constraint for a Hyperdrive pool can be formulated as:

$$
z - \frac{e}{c} \geq 0
$$

To avoid inflation attacks caused by precision errors in fixed-point arithmetic, the AMM reserves a small amount of capital to ensure its share reserves never fall below the minimum share reserves, $z_{min}$. For the same reason, a Hyperdrive pool also burns an amount of LP shares equal to the minimum share reserves when the pool is initialized. To respect the minimum share reserves, we can update our solvency constraint to:

$$
z - \frac{e}{c} \geq z_{min}
$$

Since longs and shorts can be matched up to offset the solvency constraints placed on the system, there are situations in which longs and shorts may not be able to close before maturity. This decision was made because the benefits of netting positions greatly improve the capital efficiency of the system and access to liquidity for users who are opening new positions. In economically rational markets, the risk of being unable to close a position before maturity should be low.

\subsection{Zombie Interest}

The checkpoint system applies updates to the share and base reserves that result from closing matured positions. All the proceeds owed to holders of the matured long and short positions are set aside in the zombie share reserves, $z_{zombie}$. The interest collected from these matured positions after maturity is referred to as "zombie interest." In addition to tracking any unclaimed proceeds in the share reserves, Hyperdrive also tracks the proceeds in base with the zombie base reserves quantity, $x_{zombie}$. Whenever long proceeds, $\Delta x_{long}$, and short proceeds, $\Delta x_{short}$, are set aside at maturity, $z_{zombie}$ and $x_{zombie}$ are updated in lock-step as follows:

$$
\begin{aligned}
    z_{zombie, 1} & = z_{zombie, 0} + \frac{\Delta x_{long} + \Delta x_{short}}{c} \\
    x_{zombie, 1} & = x_{zombie, 0} + \Delta x_{long} + \Delta x_{short}
\end{aligned}
$$

Long and short positions don't accrue interest after maturity. Long positions are opened with a predetermined amount of fixed interest that can be redeemed in full at maturity. Short positions have paid all of their fixed interest by maturity, so allowing them to continue to accrue variable interest would be dangerous for LPs. Whenever a long or short position is closed after maturity, the system calculates their proceeds, $\Delta x$, identically to how they would have been calculated at maturity. Instead of updating the share and bond reserves when the matured position is closed, the system updates the zombie share and base reserves as follows:

$$
\begin{aligned}
    z_{zombie, 1} = z_{zombie, 0} - \frac{\Delta x}{c} \\
    x_{zombie, 1} = x_{zombie, 0} - \Delta x
\end{aligned}
$$

Over time, the proceeds that were set aside in the zombie share and base reserves accrue interest because they are still held in the underlying yield source. The interest that has been earned, $\Delta x_{interest}$, can be measured by taking the difference between the current value of $z_{zombie}$ and $x_{zombie}$ as follows:

$$
\Delta x_{interest} = c \cdot z_{zombie} - x_{zombie}
$$

Applying this delta to the zombie share reserves results in the zombie share and base reserves being equal in value after zombie interest is collected. These collections occur during automated checkpoints and when positions are closed after maturity. Most of the interest earned on these positions is distributed to LPs, and a portion is (potentially) distributed to governance, specified by the parameter $\phi_{g, zombie}$, $0 \leq \phi_{g, zombie} < 1$. When zombie interest is collected, the share reserves are increased by the LPs' portion of the interest, and the zeta adjustment is updated by the same amount so that interest collection doesn't impact the pool's spot price or depth. The reserves are updated as follows:

$$
\begin{aligned}
    z_{zombie, 1} & = z_{zombie, 0} - \frac{\Delta x_{interest}}{c} = \frac{x_{zombie}}{c} \\
    z_1 & = z_0 + (1 - \phi_{g, zombie}) \cdot \frac{\Delta x_{interest}}{c} \\
    \zeta_1 & = \zeta_0  + (1 - \phi_{g, zombie}) \cdot \frac{\Delta x_{interest}}{c}
\end{aligned}
$$

\section{Liquidity Provision}

For longs and shorts to be opened on a Hyperdrive pool, liquidity must be provided to back the positions. Traders that add liquidity to Hyperdrive pools are called liquidity providers, or simply LPs. Instead of opening long or short positions directly, LPs take the opposing side of new trades by providing the liquidity needed to open positions. Over time, these passive traders will collect a portfolio of longs and shorts that back open positions that cannot be netted out with matching positions. In return for providing liquidity, LPs are rewarded with trading fees that offset some of the costs imposed by adverse selection.

Unlike traditional AMMs, immediate liquidity withdrawal in Hyperdrive AMMs is not always available to LPs because their liquidity may be backing open positions. To reduce the amount of monitoring needed from LPs to remove their liquidity, Hyperdrive pools allow LPs to convert their LP shares to withdrawal shares. Withdrawal shares are LP shares that can be redeemed when liquidity becomes available after some open positions are closed. Withdrawal shares are redeemed for base on a first-come-first-served basis, so it is possible for an LP that removes liquidity after another to redeem their withdrawal shares first.

Aside from the need for withdrawal shares, Hyperdrive's LP system also differs from those of more traditional AMMs in how it prices something called LP shares (covered in the next section). Since LPs own a portfolio of long and short positions,  share reserves aren't sufficient for LP share pricing. To price LP shares safely and fairly, the LP share price and core LP operations are defined from the idea that the LP share price should be invariant under all of the trading and LP operations.

\subsection{LP Shares}

The portion of a pool's liquidity owned by a given liquidity provider is tracked by the amount of LP shares they own. Liquidity providers receive LP shares when they add liquidity to a pool. These shares change in value over time as interest accrues and positions are closed. Since Hyperdrive LPs are responsible for backing the long and short positions that cannot be netted, removing liquidity from Hyperdrive is a more complicated process than removing liquidity from a conventional AMM. To offer liquidity providers fair pricing and a simple user experience, Hyperdrive's LP shares pass through three distinct phases during the LP life cycle. 

When liquidity providers add liquidity to the pool, they receive LP shares. These LP shares are considered to be active LP shares. The proceeds earned by active LP shares, from fees and closing positions, are continuously reinvested into the pool to support more trading volume. As active LPs accrue value over time, the pool becomes more liquid and supports even larger trades. The total supply of active LP shares is denoted as $l_a$. When a liquidity provider removes some of their liquidity, they receive a combination of base and withdrawal shares.

Withdrawal shares are LP shares that have been queued for withdrawal. While the liquidity underlying withdrawal shares still backs open positions, the proceeds these shares earn are not reinvested into the pool. When solvency increases after positions are closed, withdrawal shares are moved to the next phase of the LP share life cycle and marked as ready for withdrawal. The total supply of withdrawal shares, including withdrawal shares marked as ready for withdrawal, is denoted as $l_w$. Withdrawal shares that haven't been marked as ready for withdrawal are priced identically to active LP shares, but those that have been marked as ready for withdrawal are priced using special accounting rules.

When capital becomes available to buy back withdrawal shares, the pool will automatically set aside the proceeds of purchasing withdrawal shares into the \textit{withdrawal pool} and mark some withdrawal shares as ready for withdrawal. The total supply of withdrawal shares that is ready for withdrawal is denoted as $l_r$, and those that are ready for withdrawal are a subset of the total amount of withdrawal shares. Withdrawal shares can be redeemed on a first-come-first-serve basis for the proceeds in the withdrawal pool. The capital in the withdrawal pool is denoted as $z_r$, and a liquidity provider that redeems $\Delta w$ withdraw shares, $\Delta w \leq l_r$, will receive the following amount of base $\Delta x$:

$$
\Delta x = c \cdot \Delta w \cdot \frac{z_r}{l_r}
$$

For the purposes of pricing active LP shares and withdrawal shares, it is useful to define the total LP supply $l$ as the amount of LP shares that are not marked as ready for withdrawal. This quantity can be calculated as:

$$
l = l_a + l_w - l_r
$$

\subsection{LP Share Price}

Traditional AMMs price LP shares in proportion to the pool's reserves; however, this pricing model doesn't accurately price Hyperdrive LP shares because it doesn't account for the open positions owned by the LPs. To price LP shares, we calculate the \textit{LP Share Price} of the LP positions. The core insight used in Hyperdrive's LP system is that holding the LP share price invariant during instantaneous trading or the adding or removing of liquidity prevents sandwich attacks against existing LPs, and ensures that LPs are treated fairly when adding and removing liquidity. The LP share price $p_{LP}$ is calculated using the LP present value $PV$ and the LP total supply $l$, which will be defined below, and calculated as the present value per share as follows:

$$
p_{LP} = \frac{PV}{l}
$$

The LP present value is the value of the portfolio of open long and short positions owned by the pool's LPs in the current market conditions. Intuitively, this quantity is calculated by simulating the impact that closing all the open positions would have on the share reserves. Since the amount of fees traders pay depends on the spot price when their trade is executed, fees are not factored into the LP present value. Despite the exclusion of fees, the present value provides a robust valuation of the assets currently held by LPs.

To calculate the present value efficiently, we take advantage of the fact that Hyperdrive's trading function is path-independent by decomposing the outstanding positions into the net position of new bonds, $y_{new}^{net}$, and the net position of matured bonds, $y_{mature}^{net}$.  To facilitate this calculation, Hyperdrive tracks the amount of longs outstanding, $y_l$, the average maturity of long positions $t_l$, the amount of shorts outstanding $y_s$, and the average maturity of short positions, $t_s$. Using this information, we can calculate the pool's net position of new bonds, $y_{new}^{net}$, and the pool's net position of matured bonds, $y_{mature}^{net}$, as follows:

$$
\begin{aligned}
    y_{new}^{net} & = y_l \cdot t_l - y_s \cdot t_s \\
    y_{mature}^{net} & = y_l \cdot (1 - t_l) - y_s \cdot (1 - t_s)
\end{aligned}
$$

We define $n_{new}$ as the impact on the share reserves of closing the net position of new bonds. At the beginning of the term, long positions are closed by selling the underlying bonds on the trading invariant, which decreases the share reserves. In contrast, short positions are closed by purchasing the underlying bonds, which increases the share reserves. To account for the fact that some trading invariants can run out of inventory above the maximum sell amount, $y_{sell}^{max}$, or the maximum buy amount, $y_{buy}^{max}$, the part of the position that can't be sold is marked to a price of $0$ and the part that can't be bought is marked to a price of $1$. The impact of closing the net position of new bonds is calculated as follows:

$$
n_{new} = \left\{
\begin{array}{lr}
    \llap{-}I_{\Delta z}^{\Delta y} \left( 
        \min(y_{new}^{net}, y_{max}^{sell}),
        z,
        y
    \right), & \text{if } y_{new}^{net} > 0 \\
    I_{\Delta z}^{\Delta y}  \left(
        \,\,\llap{-}\min(|y_{new}^{net}|, y_{max}^{buy}),
        z,
        y
    \right) + \max(|y_{new}^{net}| - y_{max}^{buy}, 0), & \text{if }y_{new}^{net} < 0 \\
    0, & \text{if } y_{new}^{net} = 0
\end{array}
\right.
$$

We define $n_{mature}$ as the impact on the share reserves of closing the net position of matured bonds. At maturity, longs can be redeemed for their full face value, which reduces the share reserves. On the other hand, redeeming a matured short increases the share reserves by the full face value since the underlying bonds are purchased back from the pool. Since all positions can be closed at maturity, this calculation does not need to account for liquidity constraints. The impact of closing the net position of mature bonds is calculated as follows:

$$
n_{mature} = \left\{
\begin{array}{lr}
    \llap{-}M(y_{mature}^{net}), & \text{if } y_{mature}^{net} > 0 \\
    M(|y_{mature}^{net}|), & \text{if } y_{mature}^{net} < 0 \\
    0, & \text{if } y_{mature}^{net} = 0
\end{array}
\right.
$$

Now that we've defined the impact of closing the net positions of new and matured bonds, we are in a position to define the present value. This calculation simply applies the impact of closing each net position to the share reserves and subtracts $z_{min}$. This calculation simulates the total amount of liquidity that LPs could remove if all of the outstanding positions were instantly closed without considering fees. We can calculate the present value as follows:

$$
PV = z + net_{new} + net_{mature} - z_{min}
$$

\subsection{Initialization}

Hyperdrive's support for single-sided liquidity paired with virtual bond reserves requires a mechanism to adjust the number of bonds that the pool can mint when the pool is initialized. We prevent price discovery issues by mimicking the way double-sided liquidity works while holding the spot price constant. To do this, solve for the $y$ and $\zeta$ using the following systems of equations: 

\begin{align*}
& c \cdot z_e + p \cdot y = c \cdot z \\
&p = P_{spot\_price}(z_e,y)
\end{align*}
where $c \cdot z$ is the amount of base the pool is initialized with.

\subsection{Adding Liquidity}

Traders can open LP positions or add to existing LP positions by adding liquidity to the pool. When a trader adds a given amount of base $\Delta x$, the pool's share reserves are updated to include the additional base as follows:

$$
z_1 = z_0 + \frac{\Delta x}{c}.
$$

To ensure that the effective share reserves increase by the same ratio as the share reserves, the zeta adjustment is updated as follows:

$$
\zeta_1 = \zeta_0 \cdot \frac{z_1}{z_0}.
$$

The pool's spot price should remain invariant when LPs add liquidity. This can be accomplished by maintaining the same ratio between the effective share reserves and the bond reserves as we had before, which means that the bond reserves are updated as follows:

$$
y_1 = y_0 \cdot \frac{z_1}{z_0}.
$$

In return for adding liquidity to the pool, LPs are compensated with new LP shares. To ensure that LPs are fairly rewarded for adding liquidity, we solve for the amount of LP shares by preserving the LP share price as follows:

$$
\frac{PV_0}{l} = \frac{PV_1}{l + \Delta l} \implies \Delta l = \frac{(PV_1 - PV_0) \cdot l}{PV_0}.
$$

\subsection{Removing Liquidity}

When LPs remove liquidity, all of their LP shares are immediately converted into withdrawal shares. The system will attempt to redeem as many withdrawal shares as possible to return base to the LP, and any withdrawal shares that couldn't be redeemed are minted to the LP. To determine how many of the LP's withdrawal shares can be redeemed, the system first attempts to buy back as many withdrawal shares as possible using the available idle liquidity. Once the system has distributed the maximum amount of excess idle liquidity (simplified to "excess idle"), as many of the withdrawal shares as possible will be paid out from the withdrawal pool.

\subsection{Distributing Excess Idle}

The first time that excess idle is distributed is when LPs initially remove their liquidity; however, the system attempts to distribute excess idle whenever checkpoints are minted or when traders and LPs interact with the system. Distributing excess idle whenever it becomes available ensures that withdrawal shares are paid out as quickly as liquidity becomes available, which ensures that they will always be paid out within a single term after withdrawing.

The idle liquidity in Hyperdrive is defined as the amount of liquidity that can be removed by LPs without violating the solvency constraints. This idle liquidity defines an upper limit on the amount of shares that can be used to buy back withdrawal shares. We can calculate the amount of idle liquidity, $z_{idle}$, using the solvency equation as follows:

$$
z_{idle} = z - \tfrac{e}{c} - z_{min}
$$

The core idea behind the pricing of withdrawal shares is that the LP share price must be conserved whenever withdrawal shares are bought back. In order to define the problem formally, we will define $PV(\Delta z)$ to be the LP present value of the pool after $\Delta z$ shares are removed from the reserves, $l_0$ to be the starting LP total supply, and $\Delta w$ to be the amount of withdrawal shares that are paid out by the system. Using this notation, we can express the problem of determining how much idle to distribute to withdrawal shares as follows:

$$
\frac{PV(0)}{l_0} = \frac{PV(\Delta z)}{l_0 - \Delta w}
$$

Unlike the conservation of LP share price formula used when LPs add liquidity, this formula has two unknowns, $\Delta z$ and $\Delta w$. The fact that we can't just solve it algebraically means that we must find the solution by solving an optimization problem. In order to solve this undetermined system, we formulate our optimization problem to hold the starting and ending LP share prices equal as follows:

$$
\begin{aligned}
\max \quad & \Delta w \\
\text{s.t.} \quad & \frac{PV(0)}{l} = \frac{PV(\Delta z)}{l - \Delta w} \\
& \Delta z \leq z_{idle} \\
& \Delta w \leq w \\ 
& y_s \cdot t_s - y_l \cdot t_l \leq y_{out}^{max}(\Delta z)
\end{aligned}
$$

This problem can be solved efficiently using the following algorithm:

\begin{enumerate}
    \item If the pool's net curve position isn't net short, i.e. $y_l \cdot t_l \geq y_s \cdot t_s$, or if the maximum amount of bonds that can be purchased after removing all of the excess idle liquidity still exceeds the net curve position, i.e. $y_{out}^{max}(z_{idle}) \geq y_s \cdot t_s - y_l \cdot t_l$, then we set the upper bound for idle distribution to the total amount of idle liquidity, $\Delta z_{max} = z_{idle}$, and proceed directly to step (3) of the algorithm. Otherwise, we proceed to step (2). 

    \item Solve $y_{out}^{max}(\Delta z_{max}) = y_s \cdot t_s - y_l \cdot t_l$. This provides an upper bound on the amount of idle that can distributed without causing the present value to increase. The importance of this step is that it prevents pathological cases where the present value can begin to increase when capital is removed from the share reserves. These cases occur when the net curve position is short and exceeds the maximum amount of bonds that can be purchased from the pool.

    \item Solve for the amount of withdrawal shares that would be redeemed if $\Delta z_{max}$ is removed from the share reserves. This can be calculated as \\ $\Delta w = \left( 1 - \tfrac{PV(\Delta z_{max})}{PV(0)} \right) \cdot l$. If $\Delta w \leq w$, then proceed to step (5). Otherwise, set $\Delta w = w$ and continue to step (4).

    \item Solve $\frac{PV(0)}{l_0} = \frac{PV(\Delta z)}{l - \Delta w}$ for $\Delta z$ using Newton's method if $y_l \cdot t_l \neq y_s \cdot t_s$ or directly otherwise.

    \item The amount of idle that can be distributed is $\Delta z$, and the amount of withdrawal shares that the pool can buy back is $\Delta w$. The algorithm terminates.
\end{enumerate}

After the optimization problem is solved, the share reserves are debited by the amount of idle liquidity that could be distributed, $\Delta z$. The zeta adjustment and bond reserves are updated in a similar fashion to the way that they are updated when adding liquidity to ensure that the pool's spot price remains constant after distributing excess idle. The updated reserves are calculated as follows:

$$
\begin{aligned}
    z_1 = z_0 - \Delta z \\
    \zeta_1 = \zeta_0 \cdot \frac{z_1}{z_0} \\
    y_1 = y_0 \cdot \frac{z_1}{z_0}
\end{aligned}
$$

Next, the withdrawal shares that the pool can buy back, $\Delta w$, are marked as ready for withdrawal:

$$
\begin{aligned}
    l_w & \mathrel{-}= \Delta w \\
    l_r & \mathrel{+}= \Delta w
\end{aligned}
$$

Finally, idle liquidity that could be distributed, $\Delta z$, is added to the withdrawal pool:

$$
\Delta z_r \mathrel{+}= \Delta z
$$

\section*{Acknowledgements}

We are grateful for the discussions and contributions from Violet Vienhage, Charles St. Louis, Dylan Paiton, Sheng Lundquist, Giovanni Effio, Daejun Park, Tim Roughgarden, Allan Niemberg, and Ben Edgington.

\pagebreak
\section*{Disclaimer}

This paper is for general information purposes only. It does not constitute investment advice or a recommendation or solicitation to buy or sell any investment and should not be used in the evaluation of the merits of making any investment decision. It should not be relied upon for accounting, legal, or tax advice or investment recommendations. This paper reflects the current opinions of the authors and is not made on behalf of DELV or their affiliates and does not necessarily reflect the opinions of DELV, its affiliates, or individuals or entities associated with DELV. The opinions reflected herein are subject to change without being updated.

\end{document}